# Novel polystyrene-based nanocomposites by phosphorene dispersion

Elisa Passaglia[a]*, Francesca Cicogna[a], Giulia Lorenzetti[a], Stefano Legnaioli[a], Maria Caporali[b], Manuel Serrano-Ruiz[b], Andrea Ienco[b], Maurizio Peruzzini[b]

a) Istituto di Chimica dei Composti Organometallici (CNR-ICCOM), SS Pisa, Via Moruzzi 1, 56124 Pisa, Italy. E-mail: passaglia@pi.iccom.cnr.it

b) Istituto di Chimica dei Composti Organometallici (CNR-ICCOM), Via Madonna del Piano 10, 50019 Sesto Fiorentino, Italy

**Abstract**

Polystyrene-based phosphorene nanocomposites were prepared by a solvent blending procedure allowing the embedding of black phosphorus (BP) nanoflakes in the polymer matrix. Raman spectroscopy, X Ray Diffraction and TEM microscopy were employed to characterize the structural and the morphological characteristics of the achieved hybrids, with the aim to evaluate the dispersion level of black phosphorus layers. TGA, DSC analysis as well as thermal oxidation and photo-degradation techniques were employed to investigate the thermal- and the photo-stability of the samples. The collected results evidenced better thermal and photostability of both polymer matrix and dispersed layered phosphorus, suggesting really interesting polymer-nanofiller synergic effects ascribable to the presence and the good dispersion of the 2D-nanomaterial.

**Introduction**

Polymer nanocomposites developed by dispersing 2D nanostructured inorganic substrates are high performance materials exhibiting unusual property combinations and unique design possibilities.[1] The ultimate performance improvements of these hybrids systems with respect to bulk behaviour of the matrices, are intrinsically due to the nano-scaled dispersion and distribution of fillers up to the limit of an extended and active organic-inorganic interface engendering a new co-continuous nanostructured phase. Gains in mechanical and thermal behavior as well as new functional activities (gas and solvent permeability decrease, electrical and thermal conductivity, flame retardant capability, new optical features) [2-6] rely to peculiar physicochemical and structural characteristics of the nanofiller whose properties can be transferred to polymer bulk and even magnified by tuning and driving the final peculiar morphologies

Phosphorene, the monolayer of black phosphorus (BP), is recognized as a 2D material of growing interest particularly for applications in electronics and optoelectronics. BP (the most stable allotrope of elemental phosphorus) is, in fact, built up by puckered honeycomb layers of phosphorus atoms, held together via weak van der Waals forces. It can be exfoliated generating phosphorene, which exhibits a unique 2D nanostructure with peculiar anisotropy.[7] Owing to this characteristic, phosphorene possesses many intriguing properties. First, it is a semiconductor with thickness-dependent band gap: as the thickness decreases by exfoliation, the band gap gradually increases relying the nanosheets as ideal platform for electronic and optoelectronic devices.[8,9] Phosphorene shows prominent electron transport capability and low thermal conductance[10,11] both in the zig-zag and armchair lattice direction and has a really good thermal resistance,[12] resulting in a very promising material for thermoelectric applications. It is more flexible than graphene or $MoS_2$, with modulus values strongly depending on the structural anisotropy, opening the way to strain engineering applications.[13] In addition, it has been recently demonstrated to be effective as photosensitizer for singlet oxygen generation with degradation to biocompatible phosphorus oxides, highlighting its therapeutic potential in medical treatment.[14]

One of the main drawbacks in the applications of few-layer black phosphorus (2D BP) is its intrinsic instability in ambient atmosphere, undergoing to severe degradation by moisture and oxygen upon prolonged air exposition. The formation of surface oxidised species is responsible for a measurable increase in surface roughness and degradation, with severe detriment of performances of phosphorene-based electronic devices that are prepared and measured in air.[15,16] Different procedures and methodologies have been investigated to minimize the exposure to the atmosphere by encapsulating phosphorene with air-stable overlayers [17,18] or, recently, by controlling the oxidation to deliberately engineer a stable native oxide protective layer.[19]

Despite the growing interest for 2D BP and its potentiality, no work has been published regarding the preparation and characterization of nanocomposites (NCs) containing BP nanolayers dispersed in a polymer matrix, owing presumably to its intrinsic lability during the commonly employed dispersion procedures in ambient conditions.

Therefore polystyrene (PS), one of the most plastic commodity and a polymer used as model to design NCs by modulating interfacial properties and interaction with layered nanofillers,[20] is here chosen as matrix by considering the peculiar structural and electronic features of BP nanoflakes[16,21]; phosphorene is, in fact, recognized as able to interact with both electrophilic and nucleophilic molecules[22,23] and aromatic moieties[24].

After liquid exfoliation of BP with DMSO, a simple solvent blending procedure is used to provide PS-based composites with 2D BP which are structurally and thermally characterized with the dual purpose to steer the use of phosphorene in the field of polymer nanocomposites and to preserve the structure and, then, the properties of this interesting nanofiller. With these purposes, after a structural and morphological characterization of the composites aimed at assessing the presence of polymer-embedded phosphorene layers, the effect of BP nanoflakes dispersion onto the bulk properties of PS are preliminary explored by DSC, TGA and OIT measurements. At the same time the photo-stability of both the polymer matrix and the dispersed BP nanolayers in the composites are investigated by IR and Raman spectroscopy after undergoing the samples to UV-induced degradation.

**Experimental**

**Materials**

DMSO grade anhydrous (≥ 99,9 %, Aldrich) CHCl$_3$, acetone (>99%, Aldrich, HPLC grade) and MeOH (>98%, Aldrich) were used without purification.

Atactic polystyrene with Mn=86,000 D and Dispersity 2.5 (by SEC measurement, Repsol) was used without previous treatment.

**Samples preparation**

Black phosphorus (BP) was prepared by heating in a muffle oven commercially available red phosphorus together with a tin-gold alloy and catalytic amount of SnI$_4$ following a published procedure[25]. The solids were charged in a quartz tube, that was then evacuated, sealed and put in the oven and heated up to 650°C for three days. Afterwards, the cooling was very slow, with a rate of 0.1 °C /minute to afford the formation of highly pure and crystalline material (average size of the crystallites: 2 mm x 3 mm). 2D BP was prepared by liquid exfoliation following a procedure derived from literature[26]. In a typical run, microcrystalline black phosphorus (5.0 mg, 0.16 mmol) is transferred into a borosilicate tube (length = 300 mm, Øout = 15.0 mm; Øint= 11.3 mm). Dry and degassed DMSO (5.0 mL, pre-treated with molecular sieves 4Å for 24 hours, to reach a final maximum water content of 580 ppm) was added and the tube was sealed under nitrogen. The suspension was sonicated in a Elmasonic P70H sonicator at 37 KHz for 120 hours, while keeping the bath temperature at 30°C.

Two batches containing the same concentration of 2D BP (5mg/5mL) and even a low fraction of layers of BP with a thickness between 30 nm to 70 nm were used as prepared to provide the composites samples named PS1 and PS2.

To analyse the dried 2D BP by Raman, TGA and TEM, a suspension containing 5 mg of 2D-BP in 5 mL of dry DMSO was transferred in a vial and 20 mL of degassed acetone were added. The vial was put in the centrifuge for 90 minutes at 6000 rpm. Afterwards, the surnatant is removed and 15mL of degassed acetone were added to the dark grey residue. After centrifugation for 90 minutes at 6000 rpm, the surnatant was thrown away and 15mL of degassed acetone are added again. The final residue was dried under a stream of nitrogen and then under vacuum for a short time. The Raman spectroscopy confirmed the phosphorene structure while a typical TEM analysis of 2D BP showed that the single flakes and differently sized aggregates can be recognized in the dried sample [26].

The composites were prepared by a solvent intercalation procedure. Into a 250 mL two-necked round bottom flask, equipped with magnetic stirrer and previously degassed, backfilled three times with nitrogen and then left under nitrogen, 50 mL of $CHCl_3$ and 0.5 g of PS were loaded. The solution was magnetically stirred for 10 min in a continuous stream of $N_2$ until PS was completely dissolved. Under a $N_2$ current, the DMSO 2D BP suspension was (dropwise) added. The mixture was left stirring under $N_2$ for 15 min and then the mixing stopped. The flask content (a yellow/brown solution) was dropwise precipitated into 450 mL of MeOH. The polymer was then filtered, washed three times with 50mL of MeOH and dried under vacuum till constant weight. Two samples were prepared by using two batches of DMSO 2D BP suspension (samples PS1 and PS2).

For comparison purpose a blank experiment was carried out with the same methodology, without adding 2D BP suspension, providing the sample PS0.

The samples were compression molded (T = 180 °C, 10-20 Kg/cm$^2$) by using a Carver bench model 4386 to obtain the films employed for all the characterizations (with constant and uniform thickness= 90-40 μm).

Photodegradation was studied by UV radiation at different exposure times on the films PS0 and PS1 by using UV-Vis camera (UvaCube400, 400W, Hoenle) equipped with Hg lamp (high pressure Mercury Lamp with a power of 400W: emittance$_{230-285}$ = 15 mW/cm$^2$; emittance$_{330-400}$ = 11 mW/cm$^2$; emittance$_{380-500}$ = 35 mW/cm$^2$). The samples were irradiated for 40-220 min from one side.

**Characterization**

The obtained films and the UV irradiated samples PS0 and PS1 were analysed by Raman spectroscopy using a Renishaw micro-Raman Invia instrument with a 50x lens. A Nd:YAG laser at λ =532 nm wavelength was used as laser source.

Infrared spectra (FT-IR) and attenuated total reflectance (ATR) spectra of pristine and irradiated PS films were recorded by using Spectrum Two instrument of Perkin-Elmer equipped with an ATR accessory with diamond crystal. Carbonyl index (CI) was calculated after normalizing the spectra with reference to the symmetric -$CH_2$ stretching band at 2851 $cm^{-1}$. Any baseline correction was applied.

X-Ray Diffraction (XRD) patterns of PS1 composite and BP exfoliated were acquired at room temperature with a PANalytical X'PERT PRO diffractometer, employing Cu K$\alpha$ radiation ($\lambda$ = 1.54187 Å) and a parabolic MPD-mirror for Cu radiation. The diagrams were acquired in a 2$\theta$ range between 5.0° and 60.0°, using a continuous scan mode with an acquisition step size of 0.0263° and a counting time of 150 s.

TEM characterization: Images of the films were obtained using a Philips Instrument operating at an acceleration voltage 100 kV. First, a sample of PS1 film was enclosed in an epoxy resin, than the resulting disk was cut into thin lamellae which were deposited on the TEM copper grid.

Thermogravimetric (TG) analyses were carried out under nitrogen atmosphere using a Seiko EXSTAR 7200 TG/DTA instrument. TG curves were collected on samples of 1–5 mg in the temperature range from 30 to 700 °C ($N_2$ flow = 200 mL/min) with a heating rate of 10 °C/min. The onset temperature of degradation ($T_{onset}$) is defined as the intercept to the two tangents before and after the degradation step. The rate inflection temperature for the main degradation step was extracted from derivative TG (DTG) curves.

Three fractions of PS1 and PS2 samples were analysed and the data provided as average numbers with related deviations.

The glass transition temperature (Tg) of PS composites was determined with DSC using a PerkinElmer DSC4000 equipped with intracooler and interfaced with Pyris software (version 9.0.2). Calibration was performed with indium and lead as standards. The range of temperatures investigated was 40–150 °C

The oxidation induction onset temperature (OIT) was measured by heating the sample from 20 °C up to 350 °C with a heating rate of 20 °C/min and an oxygen heat flow of 60 ml/min. The samples were always heated up to a maximum temperature that was at least 20 °C higher than the steepest point of the oxidation exothermic curve[27]. The $T_{oox}$ was determined as the onset temperature of the oxidation curve outlined as the intercept point of extrapolated tangents from the baseline and the oxidation exotherm.

**Results and discussion**

The simple experimental methodology here proposed allowed producing composites (PS1 and PS2) that were structurally and thermally characterized in comparison to the blank sample (PS0). As result of solutions blending and co-precipitation approach, a grey brownish powder was obtained, which were-easily filmed by compression moulding technique; this procedure provided transparent thin films with good uniformity and reproducibility (Fig. 1a).

Raman spectroscopy, which is an effective tool for evidencing the characteristic vibrational modes of BP due to its anisotropy in the direction to the planes[28], was used to highlight the presence of BP nanolayers in the collected samples.

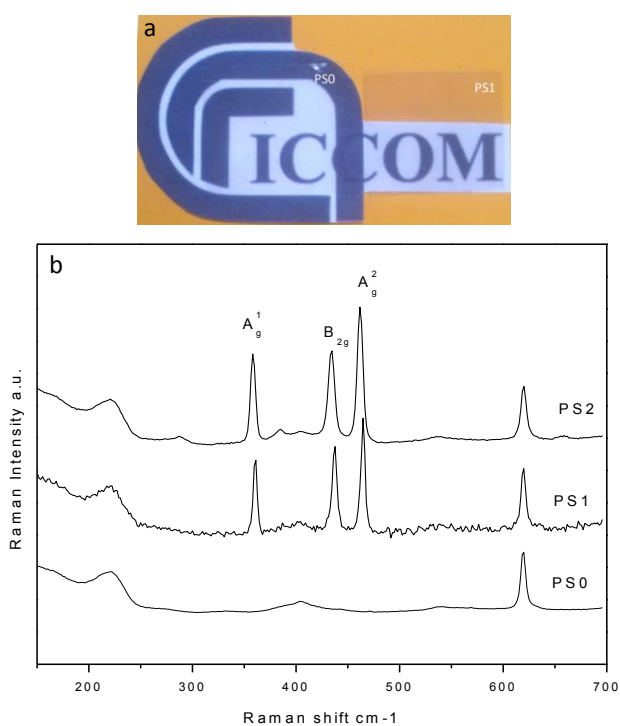

**Fig. 1** a) Photos of films obtained by compression moulding of PS samples; b) Raman spectra of PS composites (PS1 and PS2) and pristine matrix (PS0)

The three peaks at about 360, 438 and 466 cm$^{-1}$ in the Raman spectra of PS1 and PS2 samples were associated, respectively, with $A_g^1$, $B_{2g}$, $A_g^2$, Raman modes in good agreement with those already reported for BP alone indicating that BP atomic layers are crystalline[28]. These specific signals didn't change their respective intensity owing to flakes embedding in PS matrix (Fig. 1b) even if a slight blue shift was observed for PS1 sample presumably owing to the presence of thinner flake (bilayer)[16,29] The peaks at 412 and 623 cm$^{-1}$ visible in the Raman spectra of all samples can be ascribed to vibrational modes of PS matrix.[30]

XRD pattern of composite (PS1) among the broad signals due to the amorphous polymer matrix showed the diffraction peaks characteristic of exfoliated BP with high purity and crystallinity with a preferential orientation parallel to the a-b plane [31] that was maintained upon dispersion in the PS matrix (Fig. 2).

This result definitely establishes that the methodology here adopted is suitable to incorporate BP nanoflakes into PS polymer matrix. Furthermore, based on the literature, [32] a number of fractions of BP with piled phosphorene layers > 10 were evidenced, suggesting the presence in the composites of multilayer phosphorene packed nanostructures.

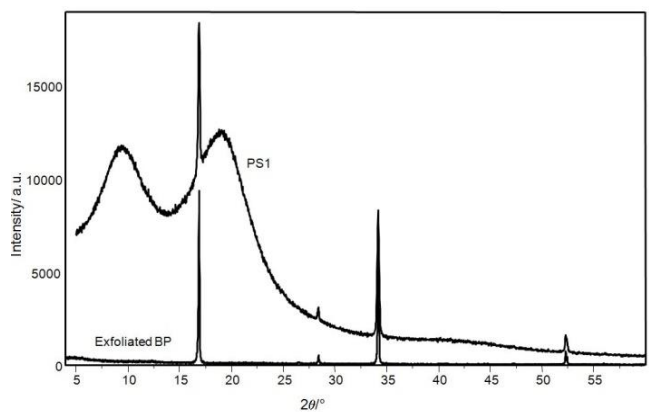

**Fig. 2** XRD diffraction paths of PS1 and starting exfoliated BP sample

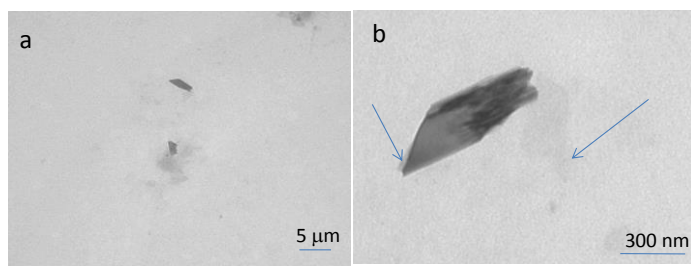

**Fig. 3** Bright-field TEM images of PS1 composite. Scale bar: 5µm and 300 nm respectively

TEM images (Fig. 3) showed mixed morphologies: single and really thin lamellae (see arrows in Fig. 3b) confirming the nanoflakes flexibility,[13] as well as piled stacks with a few layers, were easily identified. In agreement with Raman results, it was evidenced that 2D BP maintained its crystalline and oriented structure, [33] even embedded in the PS matrix.

Therefore, the procedure here adopted for composites preparation seemed to preserve the morphology of BP obtained by the liquid exfoliation methodology [26] and no larger aggregates were observed suggesting the PS macromolecules able to replace DMSO molecules without altering the pristine exfoliation degree.

The thermal transitions of the composites with reference to the blank experiment were assessed by DSC and TGA analyses (Table 1).

The two composites showed very similar features, confirming their homogeneity and the reproducibility of the preparation methodology. By comparing the data with those of blank sample an increase of Tg values was observed and attributed to the mobility hindrance of chains segment due, presumably, to the setting-up of intercalated/exfoliated nanostructures where PS macromolecules interact at interface with more rigid phosphorene layers surface and/or are sterically confined between them. The interaction between the *pz* state of phosphorene and π state of aromatic moieties of polystyrene can be reasonably suggested [34] as driving forces tuning the BP nanoflakes/aromatic rings stacking (at molecular level) as even proved by the morphological evidences.

**Table 1** Thermal transitions of BP (before exfoliation), 2D BP (after exfoliation), PS-based composites (PS1, PS2) and blank sample (PS0)

| Sample | $T_g$ [a] (°C) | $T_{onset}$ [b] (°C) | $T_{inf}$ [c] (°C) |
|---|---|---|---|
| BP | - | 481 | 524 |
| 2D BP | - | 410 | 453 |
| PS0 | 99.3 | 382±2 | 418±1 |
| PS1 | 103.2 | 409±4 | 435±2 |
| PS2 | 102.9 | 415±4 | 441±2 |

[a] calculated onto the second heating scan. [b] determined as intercept of the two tangents of degradation step [c] peak T of the DTG curved

In addition, a remarkable increase of the thermal stability of the composites samples (PS1 and PS2) was noticed: both $T_{onset}$ and $T_{inf}$ of bulk PS matrix were more than 25 degrees shifted towards higher temperature following the phosphorene dispersion, suggesting an interesting thermal stabilization effect induced by the BP nanolayers, in spite of the low filler concentration (1% wt). This behaviour, which is generally well-recognized for nanocomposites containing 2D nanofillers (cationic and anionic clays, as well as graphene) dispersed in a huge variety of polymer phases, [35-41] was here proved for the first time for composites with phosphorene. It would be presumably due to peculiar thermal conductance of 2D BP [10,11] rather than its stability to thermal decomposition[12]. The TGA analysis of both

BP and 2D BP confirmed that the thermal stability of this phosphorus allotrope is between 410°C and 480°C, higher, as expected, for the not exfoliated substrate.

However, the barrier effect to heat and gases, particularly to oxygen, associated to the 2D BP morphological shape and aspect ratio can play a significant role [42].

To deepen the evidence of thermal stabilization induced by phosphorene dispersion owing to some barrier effects, the evaluation of the oxidation onset temperature ($T_{oox}$) [27] by DSC was carried out. Films with the same thickness (90 µm) of composites and related blank sample (PS0) were analysed especially to investigate their thermal behaviour in the presence of oxygen (Fig. 4, inset table).

By heating in oxidative atmosphere, the polymer matrix and composites showed a good thermal stability till up to 220 °C. Tg is clearly higher for the PS1 sample, even in these experimental conditions.

By further raising the temperature, we collected clear evidence for the onset of the oxidation reaction temperature for both the samples: the thermal stability of the composite was significantly improved with respect to polymer matrix, and $T_{oox}$ was increased more than 20 °C, confirming the impressive stabilization effect induced by 2D BP dispersion particularly in the presence of oxygen (inset data in Fig 4).

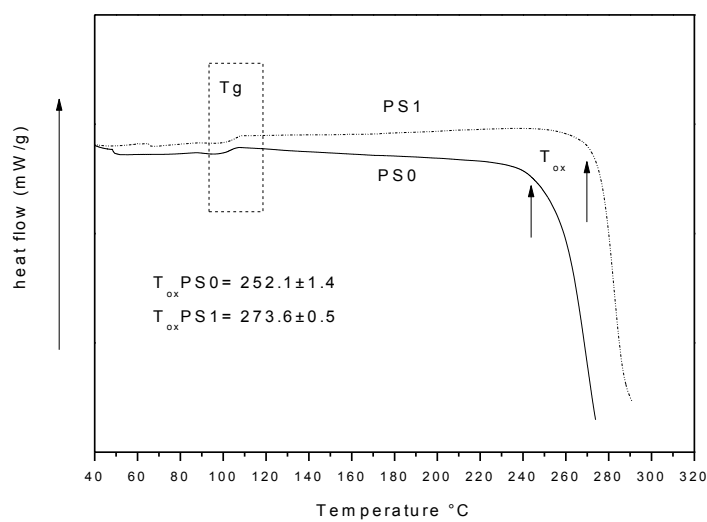

**Fig. 4** Thermoanalytical oxidation induction temperature curves of PS0 and PS1 samples and thermal parameters calculated by repeating the analysis on three different film portions ($T_{ox}$, inset data)

To better assess the composites stability in ambient conditions, the photo-degradation of the PS-based samples was studied by UV radiation: both PS0 and PS1 films were exposed to UV, simulating an accelerated light exposure in ambient conditions. The polymer samples were irradiated in air at room temperature at different times and the resulting films were analysed by ATR, FT-IR and Raman spectroscopy.

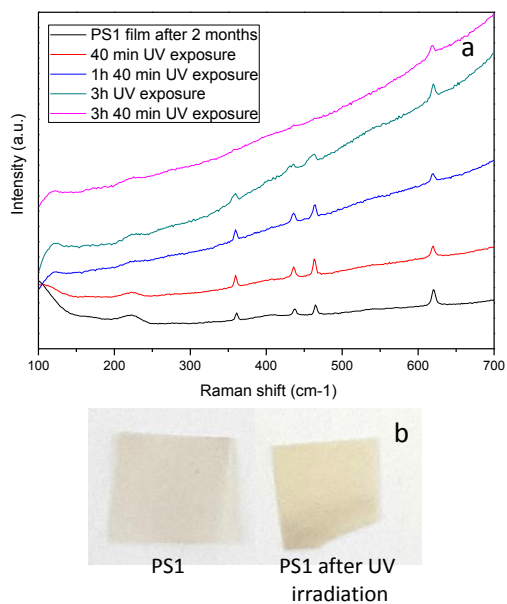

**Fig. 5** a: Raman spectra of PS1 film after 2 months of exposure to ambient atmosphere (black curve) and after irradiation with UV at different times (from red to pink curves); b: films images before and after UV irradiation

Upon two-month exposure in atmospheric environments at room temperature, the Raman spectrum of PS1 sample did not show any significant change with respect to the freshly collected sample (Fig 5a, black curve). This suggests a very good stability of the BP flakes inside the polymer bulk, at least experiencing the ambient conditions. The Raman shifts diagnostic of the BP modes were lost after 3 hours of exposure time (Fig. 5a, pink curve), when also the polymer matrix was significantly degraded.

In fact, first the toning from grey to yellow of the PS films (Fig. 5b) and, more compelling, the presence of carbonyl stretching in the FT-IR spectra for both the samples analysed (Fig. 6) confirmed the extensive oxidation of the polymer. Moreover, the matrix resulted partially insoluble to solvent used for preparing the composites and thus crosslinked, in agreement with the oxidation mechanism generally reported for polystyrene [43] as accounted also by recent literature [44]

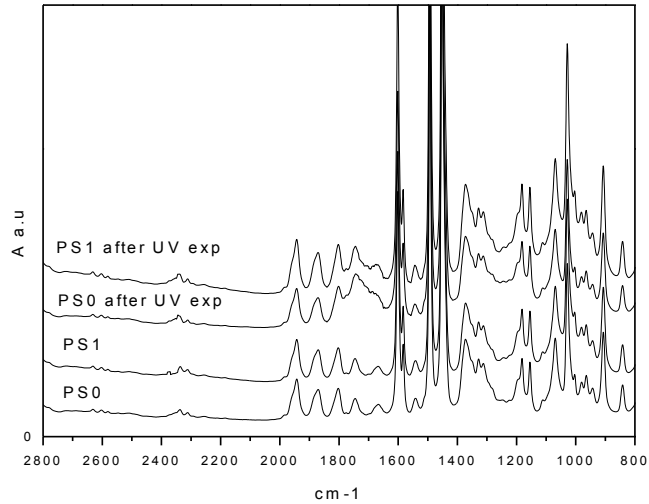

**Fig. 6** FT-IR spectra of PS0 and PS1 films before and after UV irradiation

The sample PS0 seemed extensively oxidised while fewer effects were apparently observed for PS1, whose spectrum changes are of smaller entity (see carbonyl stretching region from 2000 to 1500 cm$^{-1}$ in Fig. 6). To corroborate this observation the carbonyl index (CI) was calculated as a measure of the progress of photo-degradation in the ATR spectra on the irradiated sides of the films (Table 2, see experimental part).

**Table 2**: Carbonyl Index (C.I.) of un-treated and photo-degraded samples

| Sample | C.I.=Abs(1710 cm$^{-1}$)/Abs(2848 cm$^{-1}$) |
|---|---|
| PS0 (un)[a] | 0.38 |
| PS0 (UV)[b] | 2.37 |
| PS1 (un)[a] | 0.41 |
| PS1(UV)[b] | 1.41 |

[a] un-treated sample [b] UV irradiated for 3 hours and 40 minutes

The calculated C.I. of un-treated PS0 and PS1 films, after normalization to the reference band, showed similar values. Instead, a meaningful lower value for UV-irradiated PS1 sample was evidenced, confirming the better stability of the composite containing phosphorene even to UV radiation. Even if an easy result's rationalization may invoke the oxygen diffusion hindrance by 2D BP layers dispersed in the PS bulk, we cannot exclude that the proved UV photoresponsivity of few-layer BP [45] can play a significant role in preserving the polymer matrix, even in these conditions.

**Conclusions**

Polystyrene-based phosphorene nanocomposites were firstly prepared and characterized in terms of structural, morphological, thermal- and photo- oxidation features.

Starting from solvent dispersion of BP nanoflakes[26], these nano structured substrate was embedded in the polymer matrix by preserving the pristine exfoliated morphology, without changing the preferential orientation of platelets and aspect ratio. This evidence suggests that the electronic interactions between PS and nanolayers are able to stabilize the dispersion degree of BP. Such kind of PS-based hybrids showed improved thermal stability, even in the presence of oxidative atmosphere as well as better resistance to photo-degradation. The results can be related to the establishing of effective synergic interactions between the PS chains and the dispersed few-layers BP, suitable and able to maintain the pristine 2D BP morphological characteristics and the related barrier effects that can be preliminary invoked to explain the assessed stability to oxygen. On the other hand, and also really important, once incorporated in the polymer, BP nanoflakes showed an improved stability even when the composites were stored in air, at room temperature in ambient conditions. These observations widen the application possibilities of this really promising nanofiller once polymer-coated, particularly in making easier the design of different devices (for optoelectronic applications and/or gas/chemical sensing). Studies are in progress in our laboratories to expand the nature of polymers capable of incorporating phosphorene and to better define the bulk properties of the phosphorene-based nanocomposites.

**Acknowledgements**

The authors thank the European Research Council for funding the project PHOSFUN "Phosphorene functionalization: a new platform for advanced multifunctional materials" (Grant Agreement No. 670173) through an ERC Advanced Grant to MP.